\documentclass{IEEEcsmag}

\usepackage[colorlinks,urlcolor=blue,linkcolor=blue,citecolor=blue]{hyperref}
\expandafter\def\expandafter\UrlBreaks\expandafter{\UrlBreaks\do\/\do\*\do\-\do\~\do\'\do\"\do\-}
\usepackage{upmath,color}

\jvol{XX}
\jnum{XX}
\paper{8}
\jmonth{November}
\jname{~}
\jtitle{~}
\pubyear{2024}

\begin{document}

\sptitle{~}

\title{Engineering Trustworthy Software: A~Mission~for~LLMs}

\author{Marco Vieira}
\affil{University of North Carolina at Charlotte, Charlotte, NC, 28223, USA}

%\author{Second Author Jr.}
%\affil{Company, City, (State), Postal Code, Country}

%\author{Third Author III}
%\affil{Institute, City, (State), Postal Code, Country}

%\markboth{THEME/FEATURE/DEPARTMENT}{THEME/FEATURE/DEPARTMENT}

\markboth{~}{~}

\begin{abstract}\looseness-1LLMs are transforming software engineering by accelerating development, reducing complexity, and cutting costs. When fully integrated into the software lifecycle they will drive design, development and deployment while facilitating early bug detection, continuous improvement, and rapid resolution of critical issues. However, trustworthy LLM-driven software engineering requires addressing multiple  challenges such as accuracy, scalability, bias, and explainability.

\end{abstract}

\maketitle

\chapteri{L}arge Language Models (LLMs) are revolutionizing how we engage with technology, from improving natural language processing to automating tasks and suporting decision-making. These models are trained on extensive datasets and demonstrate great capabilities in analyzing and generating context-specific content$^{1}$.

With software deeply embedded into our daily lifes, from critical power grids and healthcare systems to financial networks and transportation, trustworthiness is becoming essential$^{2}$. A key challenge is managing complexity while maintaining quality attributes such as reliability, security, scalability, fairness, and ease of maintenance. This is increasingly difficult as software integrates a wider range of technologies (cloud computing, microservices, AI/ML components, edge devices, etc) and agile development and continuous integration and deployment (CI/CD) demand faster development cycles. 

Developers need to  deal with the balance between fostering innovation, rigorously testing code, preserving legacy systems, and addressing ethical concerns regarding privacy and AI/ML. LLMs offer a promising support to build trustworthy software systems in this context. Such models can help improve key processes across the development lifecycle, from requirements elicitation and architecture design to code generation, testing, and issue management, among others. 

In code generation, LLMs can support developers produce code with higher quality by adhering to best coding practices, in a time and cost effective manner$^{3}$. During the architecture design, LLMs can help defining secure and scalable designs that ensure systems to be resilient to threats$^{4}$. LLMs also have the potential to automate code analysis and testing by detecting code patterns and generating comprehensive test cases, minimizing the cost of fixing issues later in the process$^{5}$. Among many other examples, LLMs can also help improving issue management by analyzing bug reports, prioritizing security vulnerabilities, and supporting root cause analysis$^{6}$. 

LLMs are already being used in software development for isolated tasks, but an holistic vision that considers broader trustworthiness objectives is missing. 
Such vision consists of \textbf{integrating LLMs across the entire software development lifecycle}, from  requirements gathering to deployment of infrastructure as code (IaC) and post-deployment monitoring, allowing development teams to continuously improve their systems while considering relevant trustworthiness properties. We are, however, far from realizing this, with numerous challenges to be addressed.

Key issues include mitigating weaknesses and biases in LLM-generated code and recommendations, enhancing the explainability of decisions to build trust among developers and users, and improving the accuracy of LLM-driven assessments. Additionally, research is needed to understand how LLMs could be integrated with existing software engineering tools and practices, handle the complexity of large-scale systems, comply with legacy codebases, and ensure compatibility with standards and regulations. Until these and other challenges are resolved, the adoption of LLMs for trustworthy software remains incomplete. 

\section{TRUST AND TRUSTWORTHINESS}

Trust and trustworthiness have been extensively studied across various domains. Numerous works in the literature focus on the issue of trust in social relationships as well as trust and trustworthiness within business environments$^{7}$.

In the software context, trust can be understood as a stakeholder reliance on a system to behave as expected$^{8}$. This reliance is inherently risky, as it is often based on a subjective belief formed through past experiences with the same or other systems. Consequently, the trust level can be interpreted as the estimated probability of such reliance, which is uncertain and subject to dynamic change. In other words, trustworthiness can be defined as the degree to which a software system deserves to be trusted.

Although trust is defined differently across various fields, a common objective in all definitions is the precise assessment of trust levels$^{7}$, which serve as a foundation for informed decision-making. Thus, establishing trustworthiness is both the first and most essential step in fostering trust, calling for robust methods for design, development, deployment and assessment. %These methods enable not only the enhancement of trustworthiness but also provide a basis for comparison when required.

Trustworthiness is a key concern for developers, researchers, and enterprises$^{9}$. However, several factors contribute to the difficulty of ensuring trustworthiness. These include the diversity of software systems, the scale and complexity of modern systems, and the subjective nature of trust and trustworthiness. Depending on the context (e.g., critical or noncritical systems), different quality attributes (e.g., security or performance) may play a role in the system’s trustworthiness.

Trustworthy software requires a range of functional and non-functional requirements to be met. While the attributes and metrics used to evaluate software functionality can vary depending on the specific purpose, they remain independent of the environment in which the software operates. The relative importance of non-functional requirements, however, depends on factors such as the criticality of the software (e.g., safety-critical or business-critical), the significance of the data handled (e.g., private data), financial implications (e.g., monetary transactions), among many others. 

In short, common trustworthiness requirements include$^{9}$: security - protect data and operations from unauthorized access or breaches, ensuring confidentiality, integrity, and availability; reliability - perform functions accurately and consistently; privacy - user data must be handled responsibly, complying with regulations; robustness - handle unexpected inputs or stress conditions gracefully; maintainability - code should be designed for ease of updates and modifications; and ethical and legal compliance - adhere to ethical principles and relevant legal standards.

\section{LLMs IN TRUSTWORTHY SOFTWARE ENGINEERING}

\begin{figure*}[htbp]
    \centering
    \includegraphics[width=460pt]{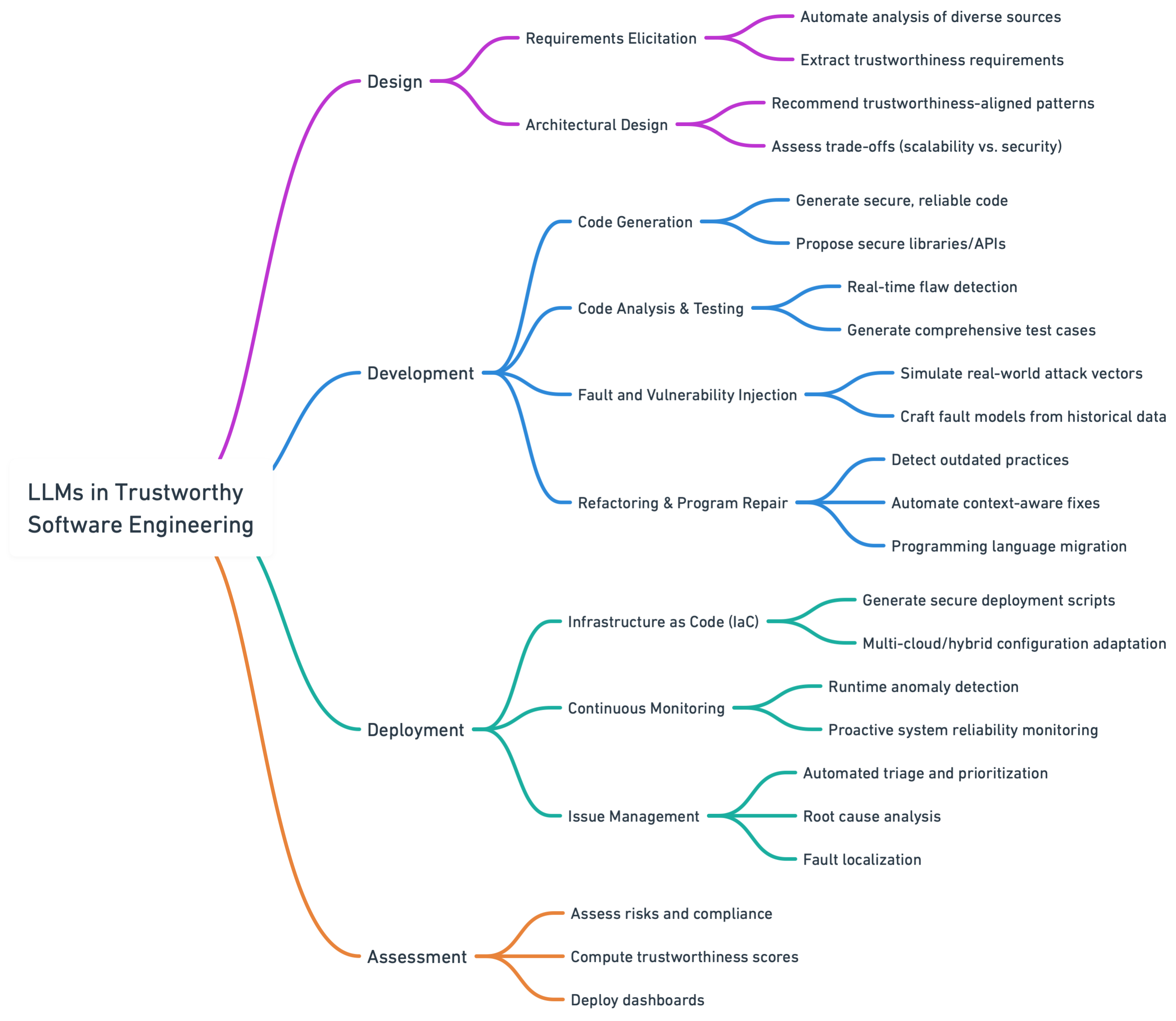}
    \caption{Role of LLMs in engineering trustworthy software.}
    \label{fig:framework}
\end{figure*}

The advent of LLMs is initiating a paradigm change in software engineering by enabling advanced solutions for tasks such as code generation, fault and vulnerability detection, and compliance assurance$^{3-6}$. However, the integration of LLMs into the software engineering lifecycle - from requirements gathering to deployment and post-deployment monitoring - calls for solutions that incorporate trustworthiness considerations from the begging. Figure~\ref{fig:framework} exemplifies how LLMs can and will play a key role in engineering trustworthy software. 

\subsection{Design}
In the \textit{requirements elicitation} phase, LLMs can facilitate a comprehensive and systematic approach to understand both functional and non-functional needs. Requirements elicitation traditionally relies on a combination of stakeholder interviews, document review, and use-case development$^{10}$. This process is time-intensive and susceptible to mistakes, particularly when it comes to non-functional requirements. LLMs can enhance this process by automating the analysis of diverse sources, including stakeholder meeting transcriptions, user stories, and regulatory documents. 

Automation can help in the identification of the relevant trustworthiness requirements, embedding security, reliability, and privacy aspects from the early stages of the development process. For instance, in applications handling sensitive personal data, LLMs can extract privacy requirements directly from regulatory frameworks like the GDPR, identifying needs for data minimization, encryption, and user consent. Similarly, LLMs can highlight security concerns by analyzing threat models. All this will contribute to a foundational stage where the principles of a secure, reliable, and resilient system are established.

Once requirements are clearly defined, the focus shifts to \textit{architectural design}, where LLMs can play a key role in guiding the creation of designs that consider and fulfill the relevant trustworthiness requirements. While architecture design typically emphasizes modularity, scalability, and maintainability, other aspects, such as resilience, fairness, data privacy, and security, are also essential to building trustworthy systems$^{11}$. LLMs can assist in this process by analyzing the requirements and recommending design patterns that align with the needs. For instance, in a distributed healthcare application where data privacy is critical, an LLM might suggest a microservices-based architecture with isolated data processing modules, reducing data exposure and improving fault tolerance.

LLMs may also be leveraged to understand the trade-offs that arise between architectural choices, such as the balance between scalability and security, enabling architects to make informed decisions. By embedding LLM-informed design principles, engineers can establish an architectural foundation based on trustworthiness attributes, thus minimizing the need for retroactive modifications to address security and reliability concerns.

\subsection{Development}
During development, LLMs can significantly improve the process of writing code that adheres to best practices and trustworthiness principles$^{3}$. The role of LLMs in \textit{code generation} should extend beyond mere syntax completion; these models should be explored to generate code that incorporates security and reliability features, ensuring that developers do not inadvertently introduce bugs and vulnerabilities. For example, LLMs assisting in the development of a web application may include input validation mechanisms to prevent common vulnerabilities such as SQL injection, while also embedding proper error-handling to ensure robustness. Furthermore, LLMs can also be leveraged to provide suggestions for integrating secure and reliable libraries or third-party APIs, which can help reduce development time and ensure trusted components to be consistently used.

The role of LLMs becomes even more pronounced in \textit{code analysis}, where exploring their capabilities for detecting flaws, vulnerabilities, and performance issues allow proactive quality assurance$^{5}$. LLMs can perform real-time code analysis by evaluating potential weaknesses as code is developed, thus identifying areas that may require rework. For instance, an LLM might detect an unprotected user input field and prompt the developer to implement validation checks. This prevents security and reliability issues from propagating through the codebase.

In addition to code analysis, LLMs can support \textit{automated testing} by generating test cases that evaluate both functional and non-functional aspects$^{5}$. By generating comprehensive test cases, LLMs can help ensuring that code is rigorously evaluated against potential edge cases, reducing the likelihood of unforeseen errors. For example, in a financial application, LLMs might generate test cases that simulate high transaction volumes to verify the system’s capacity to maintain performance. Integrating LLM-based code analysis and automated testing into CI/CD will enable continuous assessment of quality, ensuring that only code meeting trustworthiness standards is deployed.

\textit{Fault injection and vulnerability assessment} are advanced alternatives for validating error detection and intrusion tolerance mechanisms$^{12}$. LLMs can facilitate these processes by injecting faults or simulating vulnerabilities in the code, allowing engineers to observe its behavior under stress. This is particularly valuable in critical applications, where even minor faults can lead to catastrophic consequences. For example, LLMs can help creating complex attack vectors to test the system’s defensive mechanisms. 
LLMs can also be explored to define advanced fault and vulnerability models based on historical incident data from the same or other similar systems.

As codebases evolve over time, LLMs have the potentail to support \textit{refactoring} to maintain alignment with trustworthiness standards$^{13}$. For example, by detecting code patterns that may become security liabilities or performance bottlenecks, LLMs may prompt developers to revise outdated practices, keeping the codebase reliable, secure, and easy to maintain. This contributes to the long-term quality and resilience of the system, promoting trustworthiness throughout the lifecycle.

In the context of \textit{program repair}, LLMs have the potential to automatically detect and resolve defects, enhancing the trustworthiness of codebases$^{14}$. When issues arise, whether due to bugs, security vulnerabilities, or performance issues, LLMs can assist in generating context-aware fixes that align with best practices. For instance, if a null pointer exception or an unhandled edge case is detected, LLMs can be used to generate corrective code. In scenarios where a bug might be security-related, such as a missing access control check or inadequate input validation, LLMs can suggest security-enhancing code modifications to prevent exploitation. Program repair facilitated by LLMs may go beyond high-level fixes by understanding the broader implications of changes in an attempt to avoid introducing new vulnerabilities or compromise other parts of the codebase. %This not only reduces manual debugging effort but also promotes continuous code quality improvement.

\textit{Programming language migration} is often necessary to modernize legacy systems, enhance performance, improve maintainability, or adopt more secure solutions. LLMs can play a role by facilitating the process of translating code from one language to another$^{15}$. Migrating code manually is error-prone, especially in complex systems where syntax, libraries, and underlying language semantics vary significantly. LLMs, trained on diverse language pairs and programming paradigms, can help automate this process by accurately converting language-specific constructs, functions, and data structures to their equivalents in the target language. For example, migrating a system from C/C++, which often relies on pointers and manual memory management, to Rust involves adapting to a strict memory management and ownership model that emphasizes safety and concurrency. LLMs have a great potential to automate this transition.

\subsection{Deployment}
\textit{Infrastructure as Code (IaC)} plays an essential role in ensuring consistent, reliable, and secure deployments. LLMs can help generating, validating, and troubleshooting deployment configurations, reducing human error and expediting the deployment process$^{16}$. For example, LLMs might generate IaC scripts that enforce access controls, apply network segmentation, or integrate encryption protocols, aligning infrastructure configurations with security and compliance standards.
%In addition to generating configurations, LLMs analyze IaC scripts for potential security gaps, recommending adjustments to improve the system’s security posture. For instance, an LLM might detect a configuration that permits unrestricted access to a database and suggest more restrictive access policies to mitigate risk. 
This capability extends across platforms, as LLMs can translate configurations to fit multi-cloud or hybrid environments, making IaC more adaptable to varying deployment contexts. %Through the application of LLM-enhanced IaC, software deployments are made more efficient, secure, and aligned with trustworthiness requirements, reinforcing the system’s integrity from the outset.

Once the system is deployed, \textit{continuous monitoring and anomaly detection} become key to maintaining trustworthiness. LLMs can contribute to the analysis of runtime data, such as system logs and user behavior, to identify deviations from expected patterns$^{17}$. For instance, an LLM-based tool could flag unusual login patterns as a sign of unauthorized access. %By identifying anomalies early, LLMs allow operations teams to proactively address issues, preventing minor deviations from escalating into critical failures.
Through continuous monitoring, LLMs may not only support security but also enhance reliability and performance by tracking indicators such as memory usage, CPU load, and response times. This is particularly valuable in applications that require high availability, such as online transaction processing. %By maintaining vigilance over system operations, LLMs will support a dynamic trustworthiness framework that adapts to changing conditions and helps preserve stakeholder confidence.

Effective \textit{issue management} is vital for addressing incidents promptly and maintaining trustworthiness in the face of unexpected problems. LLMs will streamline issue management by automating the triage process, categorizing bugs, vulnerabilities, and incidents according to their impact and urgency$^{18}$. For example, an LLM might recognize a vulnerability affecting user authentication as critical, prompting the development team to prioritize it for immediate remediation. Automating prioritization will enable teams to allocate resources efficiently and ensure that critical issues are addressed first.

LLMs can also assist in \textit{root cause analysis} and \textit{fault localization} by detecting patterns in system logs or failure reports that reveal the underlying issues$^{19}$. In a distributed system, for instance, an LLM could analyze logs from multiple nodes to identify specific components or locations where network latency issues are impacting performance. This precise fault localization helps developers pinpointing problematic components. Additionally, LLMs can be leveraged to offer targeted remediation suggestions, providing actionable steps for developers to resolve issues quickly. This not only accelerates response times but also reduces the impact of incidents and fosters trustworthiness.

\subsection{Assessment}
Trustworthiness, as a dynamic and multifaceted quality, requires ongoing assessment to ensure that the system adapts to changing threats, user expectations, and regulatory requirements$^{9}$. In the context of software engineering, maintaining trustworthiness means that systems must not only meet initial standards but also evolve over time, while simultaneously provide trustworthiness evidences. 

Leveraging LLMs for \textit{continuous trustworthiness assessment} enables proactive evaluation against the relevant metrics both during design and at runtime.
During design, LLMs can play a crucial role in trustworthiness assessment by analyzing architectural choices and system specifications. Evaluating potential risks, such as vulnerabilities or scalability issues, and checking compliance with regulatory requirements, can provide insights that support informed decision-making. For instance, LLMs might highlight areas where additional security controls are necessary or recommend architectural adjustments to enhance resilience. At runtime, LLM-based solutions can be leverage to support trustworthiness by monitoring key events and metrics, such as security incidents or unusual user activity.

In this context, LLMs can also play a role in the real-time calculation of \textit{trustworthiness scores} and on deploying relevant dashboards that provide stakeholders with up-to-date information about the system's trustworthiness. Such scores are valuable for stakeholders who need assurance that the system consistently meets required standards in areas like security, reliability, and privacy. By continuously evaluating trustworthiness, LLMs can enable a proactive approach that informs decision-makers and supports corrective actions if trustworthiness scores indicate potential weaknesses. In practice, this continuous assessment framework provides a foundation for long-term maintenance, enabling trustworthiness to evolve in response to internal changes and external conditions.

\section{OPEN CHALLENGES}
\begin{figure*}[htbp]
    \centering
    \includegraphics[width=460pt]{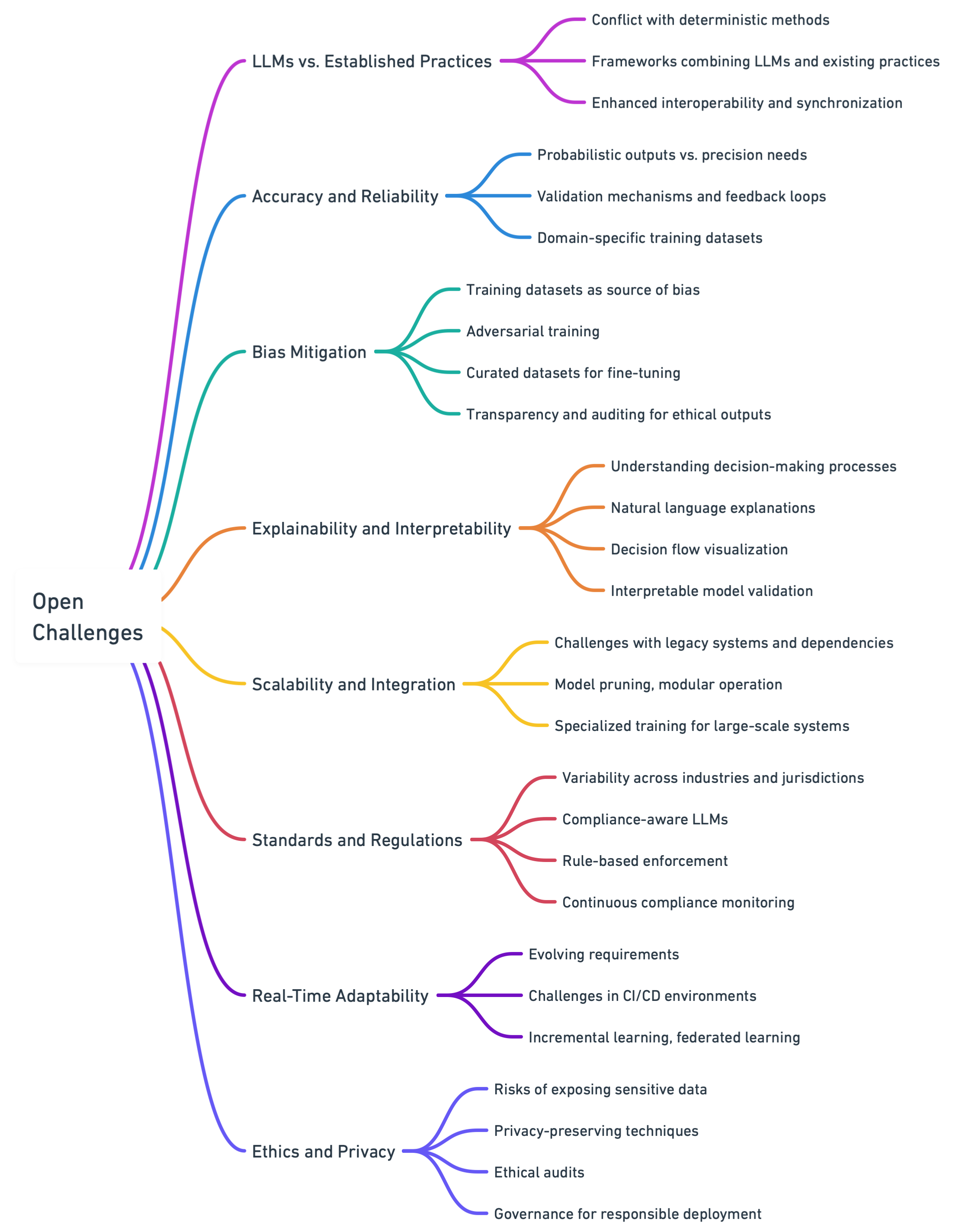}
    \caption{Challenges for LLMs in engineering trustworthy software.}
    \label{fig:challenges}
\end{figure*}

Integrating LLMs into software engineering holds great potential, but it also presents challenges that must be addressed to enable trustworthiness. In fact, to achieve a state where LLMs enhance software quality in an effective manner and without introducing new risks, research efforts must focus on issues such as integration with established software engineering practices, model accuracy, bias mitigation, decision-making explainability, and scalability, as illustrated in Figure~\ref{fig:challenges}. 

\subsection{LLMs and Established Practices}

Integrating LLMs with existing software engineering practices, techniques, and tools is a mandatory step to achieve automation, improve efficiency, and address complexity challenges. However, this integration is not simple. For example, LLM-generated outputs may conflict with deterministic methods, creating inconsistencies. Also, the reliance of LLMs on contextual data can hinder the combination with structured and systematic outputs required by traditional software engineering tools. 

Taking static code analysis for vulnerability detection as an example: traditional static analysis tools rely on precise, rule-based methods to identify specific patterns or issues in the code, producing deterministic results. If LLMs are introduced to improve this process, such as to generate explanations for detected vulnerabilities or to suggest fixes, they may produce probabilistic outputs that conflict with the findings of the static analysis tool. 

To overcome these challenges, research should focus on designing frameworks that integrate LLMs with existing techniques in a holistic manner. This includes developing hybrid approaches that combine LLM probabilistic reasoning with deterministic outputs, and fine-tuned models tailored to tools like static analysis, penetration testing, and traceability solutions. Enhanced cross-tool interoperability and mechanisms for synchronization between LLMs and traditional tools also need to be developed.

\subsection{Accuracy and Reliability}
Accuracy is foundational for LLMs in trustworthy software engineering. As LLMs are deployed across the different lifecycle stages, it becomes of critical importance to ensure that outputs are precise and free from errors. Unlike conventional automation tools, LLMs generate probabilistic outputs based on patterns learned from large datasets, which can lead to both contextually correct and misleading suggestions.
For instance, during code generation, LLMs may provide solutions that appear syntactically correct but introduce security flaws or performance bottlenecks due to insufficient contextual understanding. 

Research is needed to develop mechanisms that validate and verify LLM outputs, especially in business- and mission-critical applications where even minor issues can lead to significant consequences. Techniques such as output validation layers, specialized training datasets tailored for domain-specific contexts, and hybrid systems combining rule-based checks with LLM-driven suggestions could enhance accuracy. Additionally, integrating feedback loops, where human experts review and correct LLM outputs, may provide continuous refinement to improve accuracy and reliability.

\subsection{Bias Mitigation}
Bias in LLMs is a major challenge, as these models are trained on datasets from various sources, some containing biases related to language, coding practices, or even ethical aspects. For example, LLMs trained on open-source code might inherit biases from less secure or less efficient coding practices commonly found in certain community-contributed libraries. This can result in LLMs suggesting outdated or non-standard approaches that undermine trustworthiness.

To mitigate biases, research is needed to develop detection techniques that analyze LLM outputs for deviations from established best practices or unintended patterns. Techniques such as adversarial training, where LLMs are trained to detect and counteract their own biases, as well as fine-tuning models with curated, bias-free datasets, may help minimize these risks. Additionally, transparency in training data sources and audits of LLM-generated outputs can help identifying patterns of bias. These are essential for establishing an ethical and fair LLM integration framework that supports unbiased, trustworthy software development.

\subsection{Explainability and Interpretability}
The opaque nature of LLM decision-making processes is a significant obstacle in building trust. Developers and stakeholders need to understand the rationale behind LLM-generated outputs, especially in critical domains like healthcare, finance, and cybersecurity. However, the black-box nature of LLMs complicates interpretability, making it challenging to understand why specific architectural decisions, code patterns, or testing suggestions are recommended.

Research into explainable AI methods specifically tailored for LLMs is necessary. Explainability could involve generating natural language explanations together with the technical recommendations, or visualizing decision flows that trace how an LLM arrived at a particular suggestion. For instance, if an LLM recommends a specific architectural pattern, a corresponding explanation should indicate how this choice aligns with trustworthiness attributes like scalability and security. Developing models with built-in interpretability where simpler, interpretable models validate LLM recommendations may also help in improving transparency. 

\subsection{Scalability and Integration}
LLMs will face challenges when applied to large-scale systems, especially in complex ecosystems that rely on numerous third-party components. Modern software systems often include legacy codebases, complex interdependencies, and integration with cloud platforms, microservices, and external APIs. Embedding LLMs into these environments requires advanced research into scalability and interoperability.

Ensuring that LLMs can handle the volume and complexity of information regarding large-scale systems is crucial for a reliable integration. For example, LLMs will struggle with inconsistencies or outdated dependencies within legacy systems, resulting in erroneous recommendations or failure to adhere to best practices. Techniques like model pruning, where non-essential parameters are removed to optimize performance, or partitioning large models to operate independently across different software modules, could help address scalability issues. Additionally, specialized LLMs trained on legacy code patterns may support the ongoing maintenance of older codebases.

\subsection{Standards and Regulations}
As software systems increasingly interact with sensitive data and critical infrastructures, compliance with standards and regulations is essential for trustworthiness. However, regulatory compliance often varies across jurisdictions and industries, from GDPR in Europe to HIPAA in healthcare and PCI DSS in finance. LLMs must be able to navigate this complex landscape and generate code, recommendations, and configurations that align with the relevant regulations.

Research is needed to develop compliance-aware LLMs that can recognize regulatory requirements and integrate them into their outputs. For instance, an LLM generating infrastructure-as-code (IaC) configurations should enforce data encryption or role-based access control based on applicable regulations. Rule-based techniques should be researched to enforce regulatory constraints to be directly applied to LLM outputs. Furthermore, fine-tuning models with compliance-specific datasets could ensure that LLM-driven software engineering adheres to standards. Compliance auditing tools should also be developed to continuously monitor adherence to regulations.

\subsection{Real-Time Adaptability}
Continuous integration and continuous deployment (CI/CD) calls for adaptability in LLM-driven software engineering processes, especially as requirements evolve throughout the lifecycle. As software is deployed at rapid intervals, LLM-based solutions need to promptly provide relevant and accurate feedback. However, the latency of large models, combined with the risk of outdated recommendations as both software and requirements evolve, presents significant challenges. Also, adapting to evolving requirements is mandatory to ensure that the outputs remain aligned with existing business goals and constraints.

Research should explore LLM architectures or modular model deployment strategies that can deliver near-instant feedback. Techniques like incremental learning, where LLMs are updated continuously with recent data, can help ensuring that recommendations remain relevant. Additionally, exploring federated learning, where models are trained and deployed across decentralized locations, may help LLMs keep pace with the rapid cycles of CI/CD pipelines. 

\subsection{Ethics and Privacy}
The ethical implications of using LLMs in software engineering, particularly regarding privacy and user data, require careful consideration. Models trained on broad datasets may inadvertently expose sensitive information or generate outputs that compromise user privacy, especially if data sanitization is inadequate. Furthermore, ensuring that LLMs operate ethically is extremely important in sectors that involve sensitive applications, such as healthcare or law enforcement.

Researching the applicability of privacy-preserving techniques, such as differential privacy, to ensure that LLMs generate outputs without revealing sensitive data is mandatory. Additionally, regular ethical audits and compliance with privacy guidelines can further support the responsible deployment of LLMs. Transparent governance and ethical controls can help align LLM practices with privacy and data security values, fostering trust in LLM-driven software solutions.

\section{CONCLUSION}

The Large Language Models (LLMs) revolution in software development is just starting! LLMs are and will further transform software engineering by offering powerful tools to address the growing complexity of software systems and reduce development time and cost. The vision is clear: effectively integrate LLM-based solutions into the software development lifecycle to suport requirements elicitation, architecture design, code generation, testing, deployment, and issue management, among others. By embedding dependability and security aspects at every stage,  LLMs will enable continuous development and improvement, early detection of bugs and vulnerabilities, and faster resolution of high-risk issues, assisting developers in producing high-quality code that adheres to best practices. We are, however, very far from realizing this vision, and significant research is needed to address challenges such as the accuracy  of LLM-driven assessments,  biases in generated code and recommendations, and explainability of decisions, to ensure LLMs reach their full potential in engineering trustworthy software.

%The potential for Large Language Models (LLMs) to transform software engineering is enormous, but requires advanced research to tackle challenges like accuracy, bias mitigation, explainability, and scalability.Addressing these aspects is essential for integrating LLMs into the development lifecycle, enhancing trustworthiness in software engineering.

\section{ACKNOWLEDGMENTS}

The author acknowledges that \textit{ChatGPT 4o} has been used to improve the writing and for extracting from the text a draft version of the mind-maps presented in figures 1 and 2.

\def\refname{REFERENCES}

\vspace*{-8pt}

\begin{IEEEbiography}{Marco Vieira} is a Professor at the University of North Carolina at Charlotte (UNC Charlotte), Charlotte, NC, USA. His research interests include dependability and security assessment, fault injection, and software testing. Vieira received his Ph.D. in Informatics Engineering from the University of Coimbra, Portugal. He is Chair of the IFIP WG 10.4 on Dependable Computing and Fault Tolerance. Contact him at marco.vieira@charlotte.edu.
\vspace*{8pt}
\end{IEEEbiography}

%\begin{IEEEbiography}{Second B. Author Jr.}{\,} is a researcher at the  ABC Corporation, B\"oblingen, Germany.  Her current research interests include a, b, and c. Author received her Ph.D. degree  in physics from University. She is a Fellow of the IEEE Computer Society. Contact her at sbauthor@abc.com.\vspace*{8pt}
%\end{IEEEbiography}

%\begin{IEEEbiography}{Third C. Author III} {\,} is a program officer at the  DEF Corporation, Tokyo, Japan. Contact him at tcauthor@def.com.
%\end{IEEEbiography}

\end{document}